\documentclass[ aip, amsmath,amssymb,reprint,
]{revtex4-1}

\draft 

\usepackage{graphicx}
\usepackage{dcolumn}
\usepackage{bm}

\usepackage[utf8]{inputenc}
\usepackage[T1]{fontenc}
\usepackage{mathptmx}
\usepackage{etoolbox}
\DeclareUnicodeCharacter{2009}{\,}

\makeatletter
\def\@email#1#2{%
 \endgroup
 \patchcmd{\titleblock@produce}
  {\frontmatter@RRAPformat}
  {\frontmatter@RRAPformat{\produce@RRAP{*#1\href{mailto:#2}{#2}}}\frontmatter@RRAPformat}
  {}{}
}%
\makeatother

\begin{document}


\title{THz optical solitons from dispersion-compensated antenna-coupled planarized ring quantum cascade lasers}
\author{Paolo Micheletti}
    \affiliation{Institute for Quantum Electronics, ETH Zurich, 8093 Z\"urich, Switzerland}%
    \email{pmicheletti@ethz.ch}
\author{Urban Senica}
    \affiliation{Institute for Quantum Electronics, ETH Zurich, 8093 Z\"urich, Switzerland}%
\author{Andres Forrer}%
    \affiliation{Institute for Quantum Electronics, ETH Zurich, 8093 Z\"urich, Switzerland}%
\author{Sara Cibella}%
    \affiliation{CNR-Istituto di Fotonica e Nanotecnologie, Rome, Italy }%
\author{Guido Torrioli}%
    \affiliation{CNR-Istituto di Fotonica e Nanotecnologie, Rome, Italy }%
\author{Martin Franki{\'e}}
    \affiliation{Institute for Quantum Electronics, ETH Zurich, 8093 Z\"urich, Switzerland}%
\author{J{\'e}r{\^o}me Faist}
    \affiliation{Institute for Quantum Electronics, ETH Zurich, 8093 Z\"urich, Switzerland}%
\author{Mattias Beck}
    \affiliation{Institute for Quantum Electronics, ETH Zurich, 8093 Z\"urich, Switzerland}%
\author{Giacomo Scalari}
    \affiliation{Institute for Quantum Electronics, ETH Zurich, 8093 Z\"urich, Switzerland}%

\date{\today}

\begin{abstract}
 \textbf{Quantum Cascade Lasers (QCL) constitute an intriguing opportunity for the  production of  on-chip optical Dissipative Kerr Solitons (DKS): self-organized optical waves which can travel while preserving their shape thanks to the interplay between Kerr effect and dispersion. Originally demonstrated in passive microresonators, DKS were recently observed in mid-IR ring QCL paving the way for their achievement even at longer wavelengths. To this end we realized defect-less THz ring QCLs featuring anomalous dispersion leveraging on a technological platform based on waveguide planarization. A concentric coupled-waveguide approach is implemented for dispersion compensation whilst a passive broadband bullseye antenna improves the device power extraction and far field. In these devices, comb spectra featuring sech$^2$ envelopes are presented for free-running operation. This first hint of the presence of solitons is further supported by the observation of highly hysteretic behaviour and by phase-sensitive measurements which show the presence of self-starting 12 ps-long pulses in the reconstructed time profile of the emission intensity. These observations are in very good agreement with our numeric simulations based on a Complex Ginzburg-Landau equation time-domain solver. Such devices constitute a new experimental platform for the study of soliton phenomena in the THz range, allowing as well on-chip, passive ultrashort THz pulse generation appealing for a variety of applications.  }
\end{abstract}


\maketitle 

\section{Introduction}

\begin{figure*}[!htb]
	\centering
	\includegraphics[width=\linewidth]{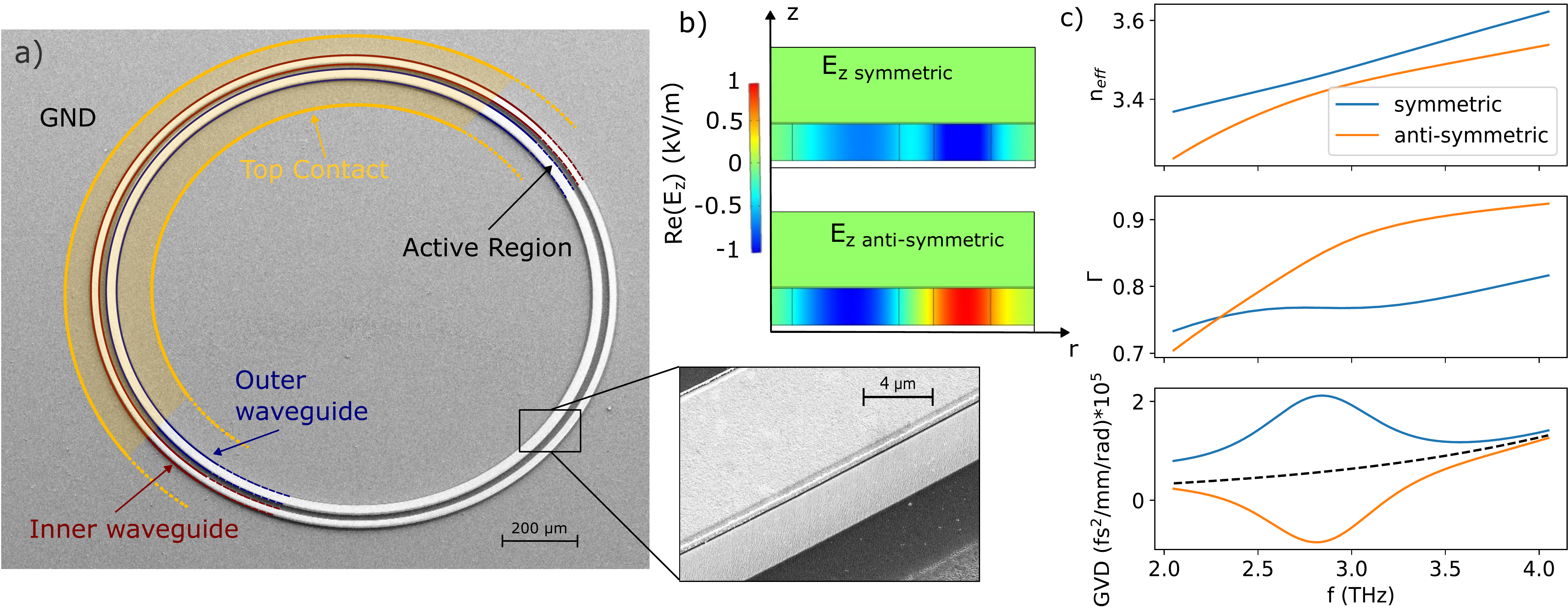}
	\caption{a) SEM image of a 650 $\mu$m radius double waveguide ring before BCB planarization. A schematic representation of the Top contact (yellow) is reported on the picture. The inner and outer waveguides are indicated in red and blue respectively. b) Vertical component of the E field of the symmetric (top) and (anti-symmetric) supermode in a double waveguide. c) Simulated effective refractive index, overlap factor ($\Gamma$) and GVD.}
	\label{fig:intro}
\end{figure*}

Recent years have seen the rapid development of passive ultrahigh-Q factor chip-scale micro-resonators as effective tools to produce Dissipative Kerr Solitons (DKS): self-organised optical waves able to propagate in a dispersive, lossy and non-linear medium while maintaining their shape and amplitude\cite{kippenberg_dissipative_2018,herr2014temporal,matsko2011mode}. These waves can exist in a medium where the spreading of a wave packet due to the dispersion is balanced by the self-modulation induced by the Kerr non-linearity, and where losses are compensated by the injected optical power. Besides their purely academical interest, these devices have already proven to be an attractive technology for an extremely wide variety of applications such as coherent telecommunication\cite{marin2017microresonator}, ultrafast optical ranging\cite{trocha2018ultrafast}, precision metrology\cite{picque2019frequency, yang2017microresonator} and frequency synthesis\cite{spencer2018optical,weng2020frequency}. 
Recently, Quantum Cascade Lasers (QCL) are emerging as promising alternative to obtain solitons in the Mid-IR range. The intersubband transition in quantum cascade heterostructures naturally provides both for a giant Kerr non-linearity\cite{friedli2013four, opavcak2021frequency} and optical gain, thus making QCLs an ideal platform for DKS. Using an active medium in particular avoids the need of an external pump \cite{scalari_-chip_2019} to sustain the parametric oscillation in the cavity, while relaxing the requirement of ultrahigh-Q factors at the same time. Nevertheless, while QCL frequency comb operation has been demonstrated in the mid-IR\cite{hugi2012mid} and THz\cite{burghoff2014terahertz} frequency domains in the past years, their operation as DKS was reported in mid-IR only very recently \cite{meng2022dissipative}. Quantum cascade laser frequency combs in free-running standard Fabry-Perot QCLs are indeed typically characterized by a quasi-continuous waveform combined with a linear\cite{singleton2018evidence,hillbrand_coherent_2019} or more complex\cite{burghoff2015evaluating} frequency modulation in the mid-IR and THz domain respectively. The reason behind this behaviour is the extremely fast dynamics ($\sim$ ps timescale) of the saturable gain combined with the spatial hole burning (SHB) arising from the presence of standing waves\cite{agrawal1988population, piccardo_harmonic_2018, opavcak2019theory, li2022real} and the presence of non linearities \cite{Burghoff:2020uh}. This issue can be overcome by exploiting travelling-wave resonators, such as ring cavities which can sustain whispering gallery modes exhibiting a well defined direction of propagation. In this case the absence of spatial hole burning, in combination with an anomalous dispersion (GVD $<0$), led to the demonstration of solitons in mid-IR QCL\cite{meng2022dissipative}.
Following this success, recently there have been demonstrations of THz ring QCLs\cite{jadida2021comb,jaidl2022silicon} frequency combs. Nevertheless achieving DKS in ring THz QCLs still presents some unsolved challenges. First of all a negative dispersion is necessary to balance out the Kerr non-linearity of the active medium. Dispersion compensation structures should therefore be exploited to correct for the large positive GVD of III-V heterostructures at THz frequencies ($\simeq10^5 \textrm{ fs}^2/mm$ ), mainly caused by the nearby optical phonon resonance (GaAs restrstrahlen band extends from 8 to 10 THz). A more practical issue concerns the power extraction. While defect-free rings are necessary to suppress SHB, the absence of a preferential site for light extraction leads to a poor output power and far-field. This issue is particularly critical for applications and, in a first place, for phase sensitive measurements, which are fundamental to fully characterize the comb state. Those indeed require fast THz detectors which typically present a limited responsivity at high frequencies\cite{sizov2018terahertz}. In the present work we intend to address these challenges, leveraging on a high-performance planarized double-metal waveguide platform\cite{senica2022planarized}. This technique is based on a standard double metal waveguide\cite{Williams:2007p107} which is encapsulated in a Benzocyclobutene (BCB) layer. Such planarization enables the disentangling of the laser cavity design with respect to the current injection pad and the RF coupling. This allows to explore alternative top contact designs allowing the fabrication of ultra-thin ring cavities, coupled double ring waveguides for dispersion compensation and to integrate passive metallic structures onto the polymer for light extraction and far field engineering.

\section{Simulations and Design}
\label{section:sym}

\begin{figure*}
	\centering
	\includegraphics[width=\linewidth]{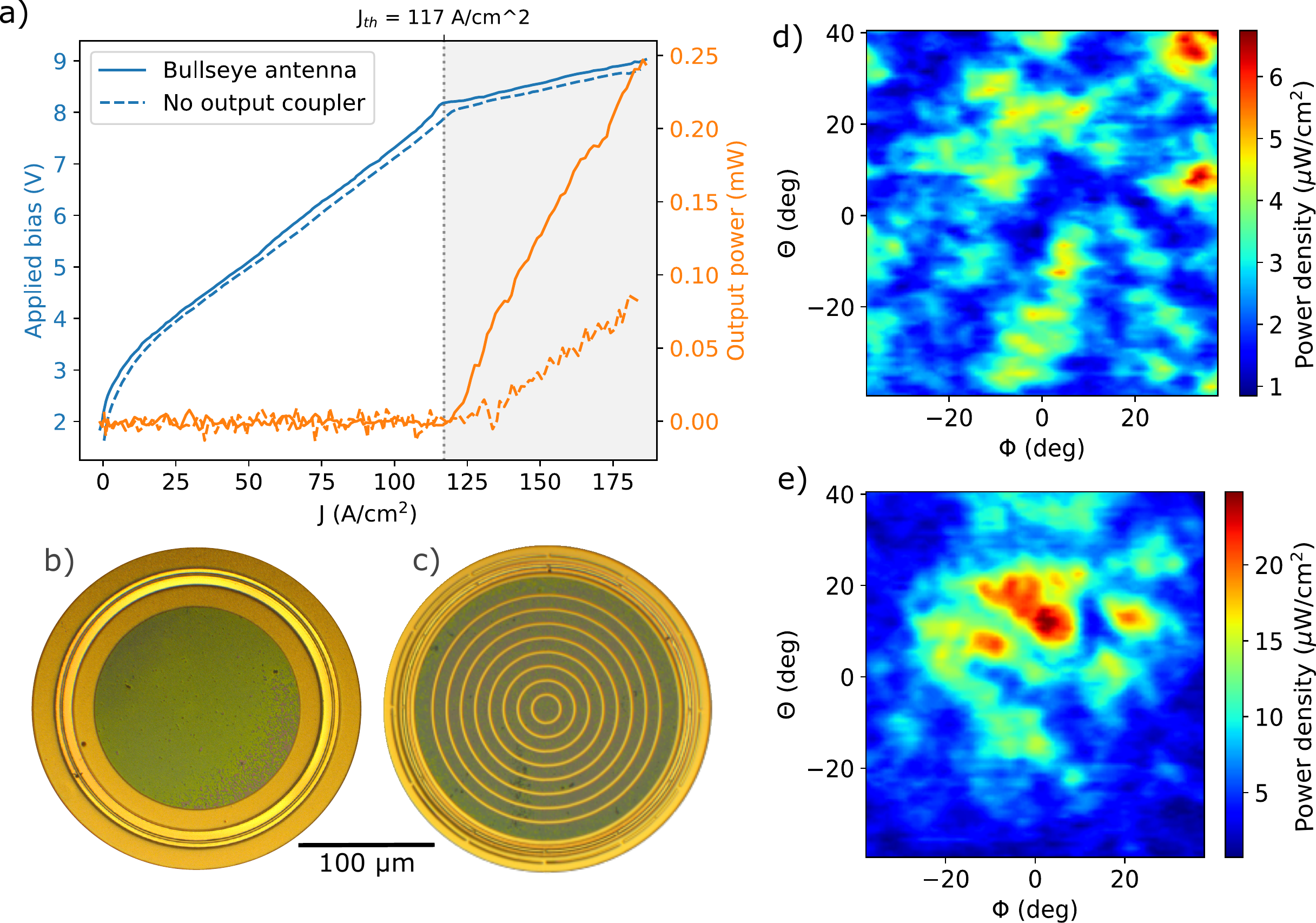}
	\caption{a) Voltage and output power as a function of current density of two devices with (solid lines) and without (dashed lines) bullseye antenna. The typical threshold is indicated with the dotted line, the gray area represent the region above threshold. Optical microscopy of a fully processed device with simple top contact (b) and integrated bullseye antenna (c). Far-field measurement of a ring without d) and with e) antenna.}
	\label{fig:power}
\end{figure*}

A double waveguide structure is considered which consists in a couple of concentric ring waveguides separated by a thin gap (Fig. \ref{fig:intro} a). Being perfectly axi-symmetric, this structure provides for the anomalous dispersion without introducing any source of back scattering and, in contrast to chirped mirrors, is compatible with integrated ring cavities. The two waveguides are designed performing 2D simulations using COMSOL Multiphysics$^{\textregistered}$. The dimensions are chosen such that the propagation vectors of the modes in each waveguide are equal for a given resonant frequency. If the gap between the waveguides is narrow enough (in practical terms $\lesssim$10 $ \mu$m) to allow an overlap between the two modes, these can couple, forming a symmetric and anti-symmetric supermode where the global GVD is respectively enhanced or suppressed \cite{bidaux2018coupled}. Having a node in between the waveguides, the anti-symmetric supermode features a higher overlap factor $\Gamma$ (Fig. \ref{fig:intro} b-c), hence a lower lasing threshold. Therefore the laser would naturally select the mode which exhibits negative GVD.
\begin{figure}
	\centering
	\includegraphics[width=1\linewidth]{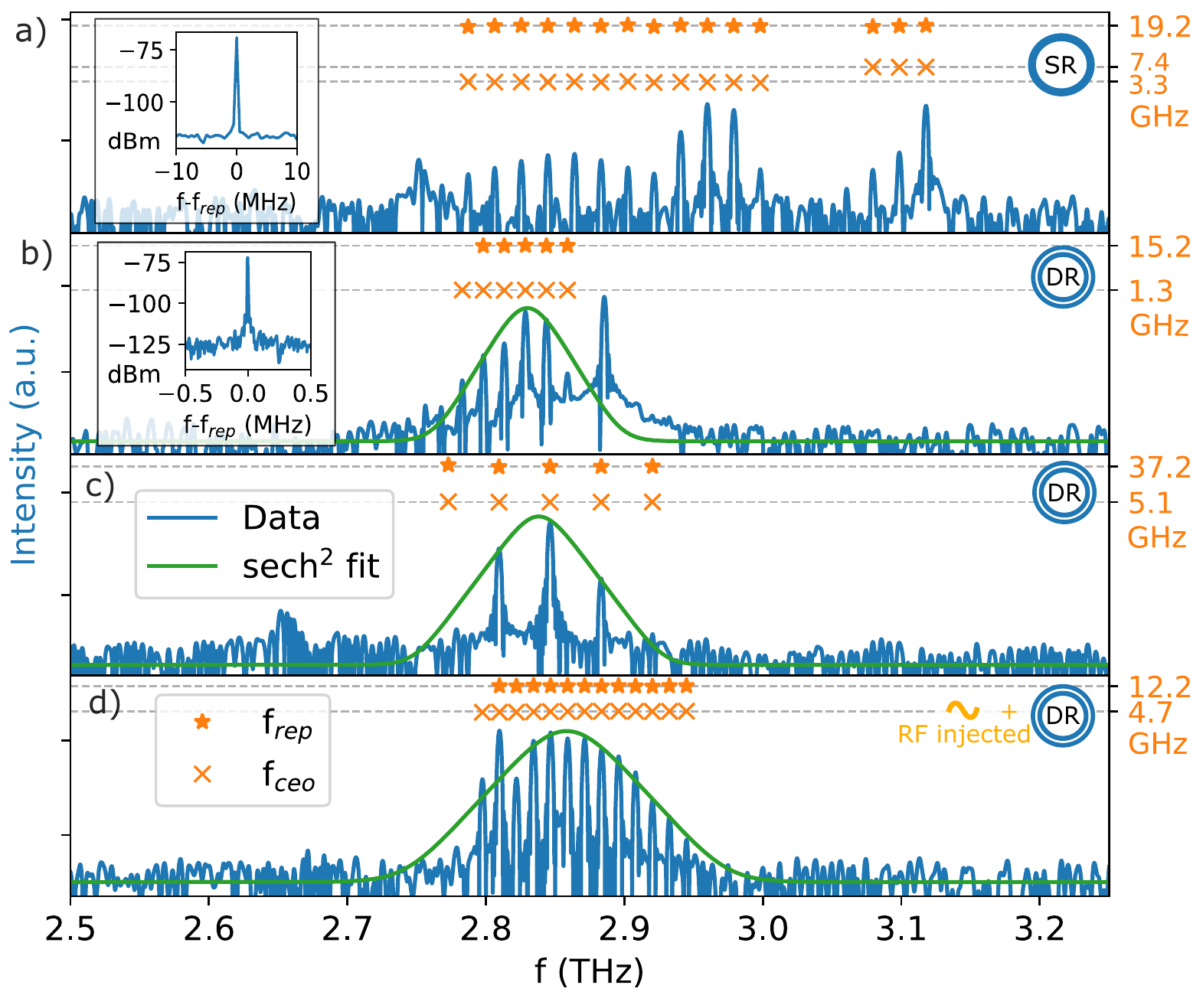}
	\caption{CW spectra of different devices (a-d blue traces) are displayed in log-scale, with a sech$^2$ fit of the spectral envelope (b,c,d green trace). The peak spacing, i.e. f$_{rep}$ ($\star$) and the f$_{ceo}$ ($\times$) computed from the DC spectra are reported on top. a) Free running spectrum of a single ring (SR) with a 650 $\mu$m radius and a 60 $\mu$m wide waveguide. The device is driven in CW at 750 mA (9.2 V) and kept at a heat sink temperature of 20K. The device beatnote measured from the bias-tee with a Rohde\&Schwarz FSW-67 spectrum analyzer is shown in the inset. b)  Free running spectrum of a 800 $\mu$m radius double ring (DR) with waveguide widths of 35 $\mu$m (Inner) and 25 $\mu$m (Outer) biased with 1.045 A (9.5 V) at 35K. The inset reports an electrical measurement of the beatnote. c) Free running spectrum of a 1 mm radius DR ring. The waveguides width is 36.5 (Inner) and 28.5 (outer) $\mu$m. The laser is biased with 0.96 A (8.8 V) at 30 K. d) The same device is strongly injected with a +32 dBm RF signal with a frequency f$_{inj}$ = 12.35 GHz while keeping the device at 30K and with a 0.8 A (8.5V) bias.  }
	\label{fig:spec3}
\end{figure}
As previously mentioned this design was enabled by a planarized waveguide platform which allows to fabricate waveguides below 40$\mu$m in width without directly bonding on the heterostructure. For this work we used a low-threshold (J$_{thresh}\leq 140$ A/cm$^2$), broadband active region based on a strongly diagonal transition GaAs/AlGaAs homogeneous heterostructure \cite{forrer_photon-driven_2020}. After the definition of the waveguides with a self-aligned dry etching process (Fig. \ref{fig:intro} a) a thick layer of BCB is spun and baked. The polymer is then etched down with RIE at the same level of the waveguide top metalization, which is exposed, and a wider metal contact is deposited on top (Fig. \ref{fig:power} b). In this way the device can be wirebonded on top of the polymer avoiding any defect that could be introduced by bonding directly on top of the active region. Furthermore a passive bullseye antenna\cite{scheuer2003annular} can be integrated on top of the polymer improving the output power and the farfield of the device ([S2], Fig. \ref{fig:power} c). Again the choice of this solution was driven by the necessity of keeping a strictly axi-symmetric geometry to avoid backscattering. At the same time, even though a good fraction of the scattered THz radiation is emitted radially outwards, keeping the antenna inside the ring helps to minimize the surface occupied by the device. This constitutes indeed a practical issue since the requirement of a repetition rate of the order of $\sim$20 GHz imposes a relatively large minimum ring radius (R $\geqslant$ $\sim$650 $\mu$m).

\section{Results and Discussion}

\begin{figure*}
	\centering
	\includegraphics[width=\linewidth]{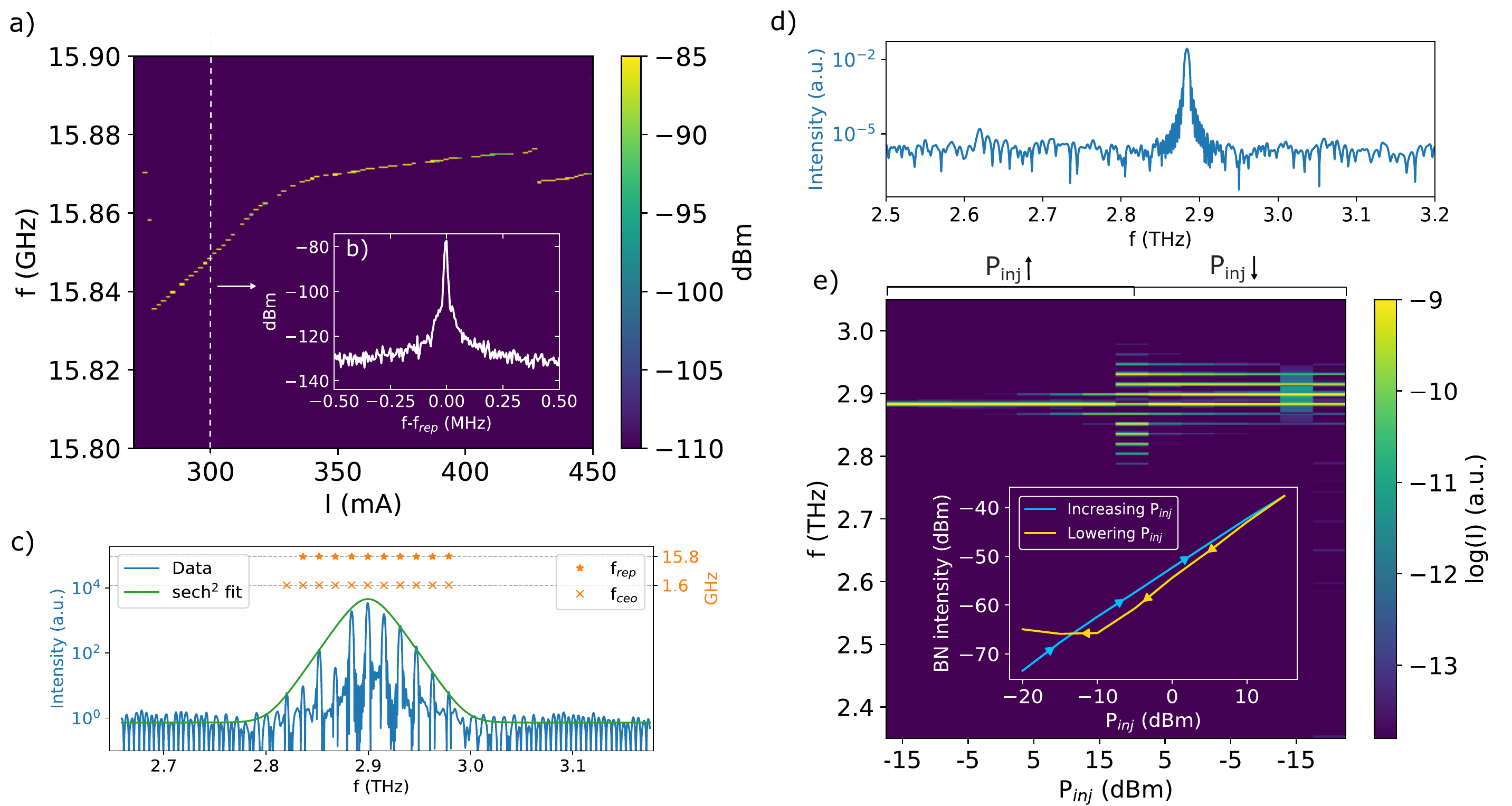}
	\caption{a) Electrical beatnote map of a 800 $\mu$m radius and waveguide width of 22.8 and  17.2 $\mu$m (inner and outer respectively) antenna-coupled ring laser measured from the device bias tee. The laser is driven in CW at 20K. The beatnote (b) and spectrum (c) correspond to a driving current of 300 mA (8.5 V). d) Spectrum of the same device and in the same operation condition as in c) but reaching the bias point increasing the current slowly. e) THz spectra as a function of RF injection power. The bias point and temperature are the same as in c) and d). The intensity of the electrically measured beatnote during the power cycle is reported in the inset.  }
	\label{fig:hyst}
\end{figure*}

We report comb operation in single- and double-waveguide free-running ring lasers (Fig. \ref{fig:spec3}). In single ring (SR) devices, i.e. in  absence of dispersion compensation, spectra with a bandwidth of $\sim$ 400 GHz are observed in Continuous Wave (CW) operation featuring a single beatnote with linewidth $<$ 10 kHz, probably limited by technical noise (Fig. \ref{fig:spec3} a). Nevertheless the spectrum does not present an hyperbolic secant (sech$^2$) envelope which is expected for solitons. Furthermore, although the mode spacing appears constant across the whole spectrum, the f$_\mathrm{ceo}$ extracted from the FTIR DC-interferogram measurement indicate the presence of two distinct sub-combs\cite{forrer2022spectral}. In contrast, a very different behaviour is observed in a dispersion-compensated ring laser (DR) (Fig. \ref{fig:spec3} b). Here, despite the lower bandwidth, the free-running CW spectrum presents a well defined sech$^2$ envelope that hints at the presence of DKS. The mode spacings and f$_\mathrm{ceo}$ appear constant across the spectrum with the exception of the highest frequency mode which can be attributed to the symmetric transverse waveguide super-mode. A further improvement can be obtained by RF injection-locking the QCL at a frequency close to the repetition rate. In Figure \ref{fig:spec3} (c-d) we report the results for a 1 mm-radius, dispersion-compensated ring laser. In free-running, the device behaves as an harmonic comb\cite{kazakov_self-starting_2017,ForrerAPL2021Harmonic} for most of its operating range (Fig. \ref{fig:spec3} c). This state also features a sech$^2$ envelope but with only few modes due to the high spacing frequency (3$\times$f$_{rep}$). When RF injected around the repetition rate with a high power microwave tone (+32 dBm at the source) the device enters in a dense comb regime with a bandwidth of $\sim$ 200 GHz and a clear sech$^2$ spectral envelope. Although it is not possible to directly measure the electrical beat-note due to the strong RF injection, the symmetry of the interferogram and the invariance of both mode spacing and f$_{ceo}$ throughout the spectrum are good indication of the comb state coherence [S4]. 

Nevertheless a measurement of the phases of the modes is necessary to prove their coherence and to reconstruct the time domain profile of the THz radiation. This is particularly challenging with the devices presented so far since their output power is usually of the order of $\sim$ 100 $\mu$W and consist of radiation scattered in every direction in-and out of plane. Hence the far-field of these devices is typically very patterned over a wide angular spread and it is difficult to effectively collect all the output power (Fig. \ref{fig:power} d). This issue can be solved with the introduction of a bullseye antenna, whose design is described in the previous section. The surface emission of two devices with and without antenna is presented in Figure \ref{fig:power} a) which indicates an increase of emitted power of a factor of $\sim$ 2.5. However, since the two devices also feature a different active area, normalizing the output power with respect to the active area returns an increase of effective power of a factor 4.4. Furthermore an improvement in the device far-field due to the antenna can be clearly seen comparing Figure \ref{fig:power} d) and e). 

\begin{figure*}
\centering
\includegraphics[width=\linewidth]{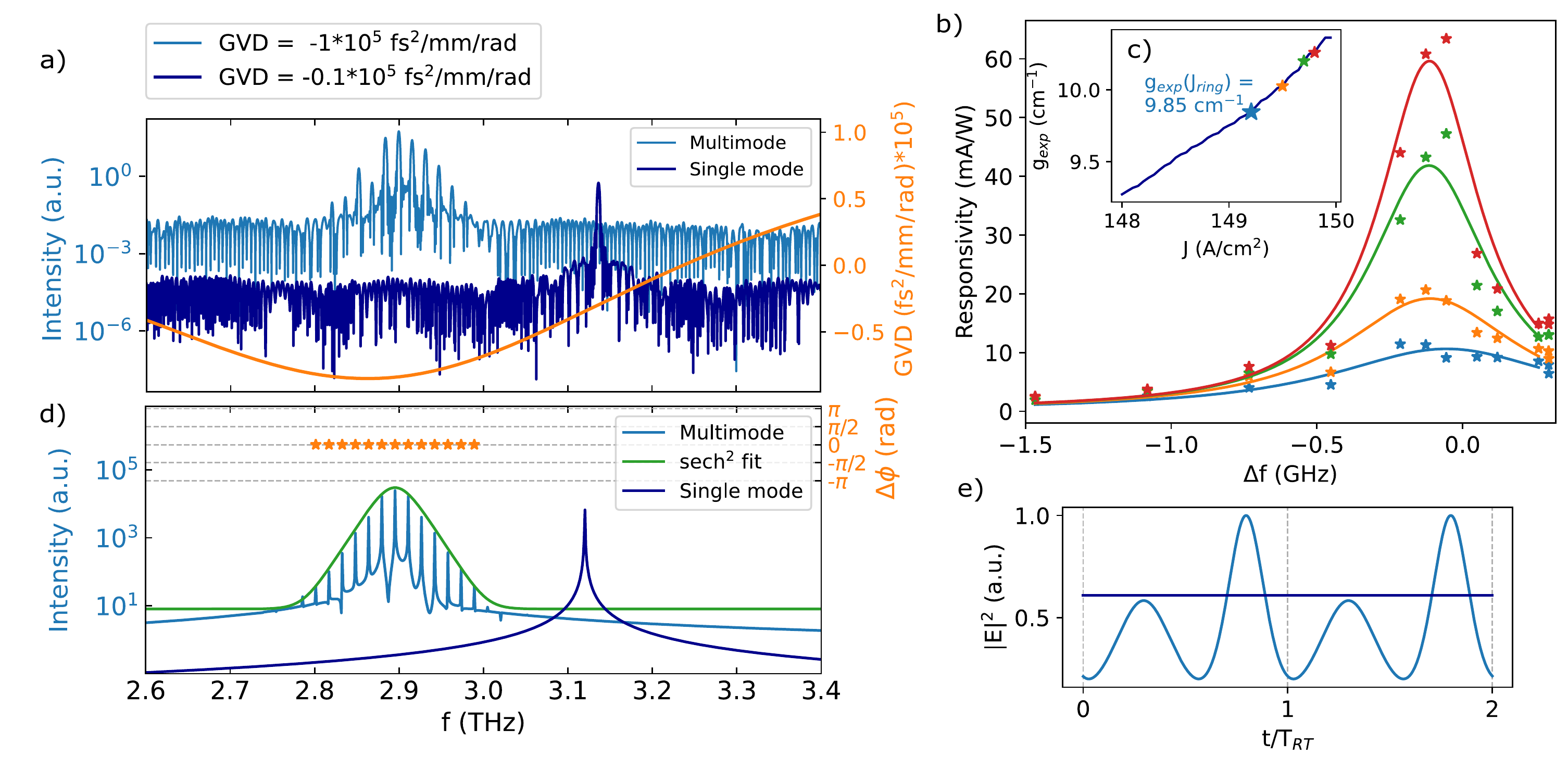}
\caption{Comparison between the spectrum of the device presented in Fig. \ref{fig:hyst} c) (light blue) and a device with same nominal dimensions (darkblue trace) operating in single mode while biased at 300 mA (8.5 V) a). The dispersion corresponding to the design of the devices obtained in the simulation described in section \ref{section:sym} is superimposed on the spectra (orange trace). b) Responsivity of a 2mm long, 50 $\mu$m wide ridge laser biased below threshold as a function of the frequency detuning with respect to the source laser for different bias currents. The experimental data ($\star$) are fitted with the model published in Ref. [35]. In the fit (solid lines) we assumed an in-coupling efficiency $\eta_{opt}$ = 1$\cdot$10$^{-4}$, $\alpha_{w}$ = 7 cm$^{-1}$ and front and back facet reflectivity of 0.4 and 0.6 respectively. The front facet reflectivity being lower due to the presence of the silicon lens. The gain extracted by fitting the responsivity curves is reported in (c) as a function of current density. The points relative to the data shown in (b) are indicated with the $\star$. The light-blue one in particular indicates the current density relative to the spectra reported in a). Simulated spectra obtained solving the CGLE equation considering GVD = -1$\cdot$10$^5$ (light-blue spectrum) and -0.1$\cdot$10$^5$ rad fs$^2$/mm (dark-blue spectrum) d). The phase difference of the modes (orange $\star$) and a sech$^2$ fit of the spectral envelope (green trace) are reported for the multimode case. The field intensities relative to the two simulated states are displayed in c) as a function of time, normalized on the round-trip period T$_{RT}$.}
\label{fig:CGLE}
\end{figure*}

\begin{figure*}
	\centering
	\includegraphics[width=\linewidth]{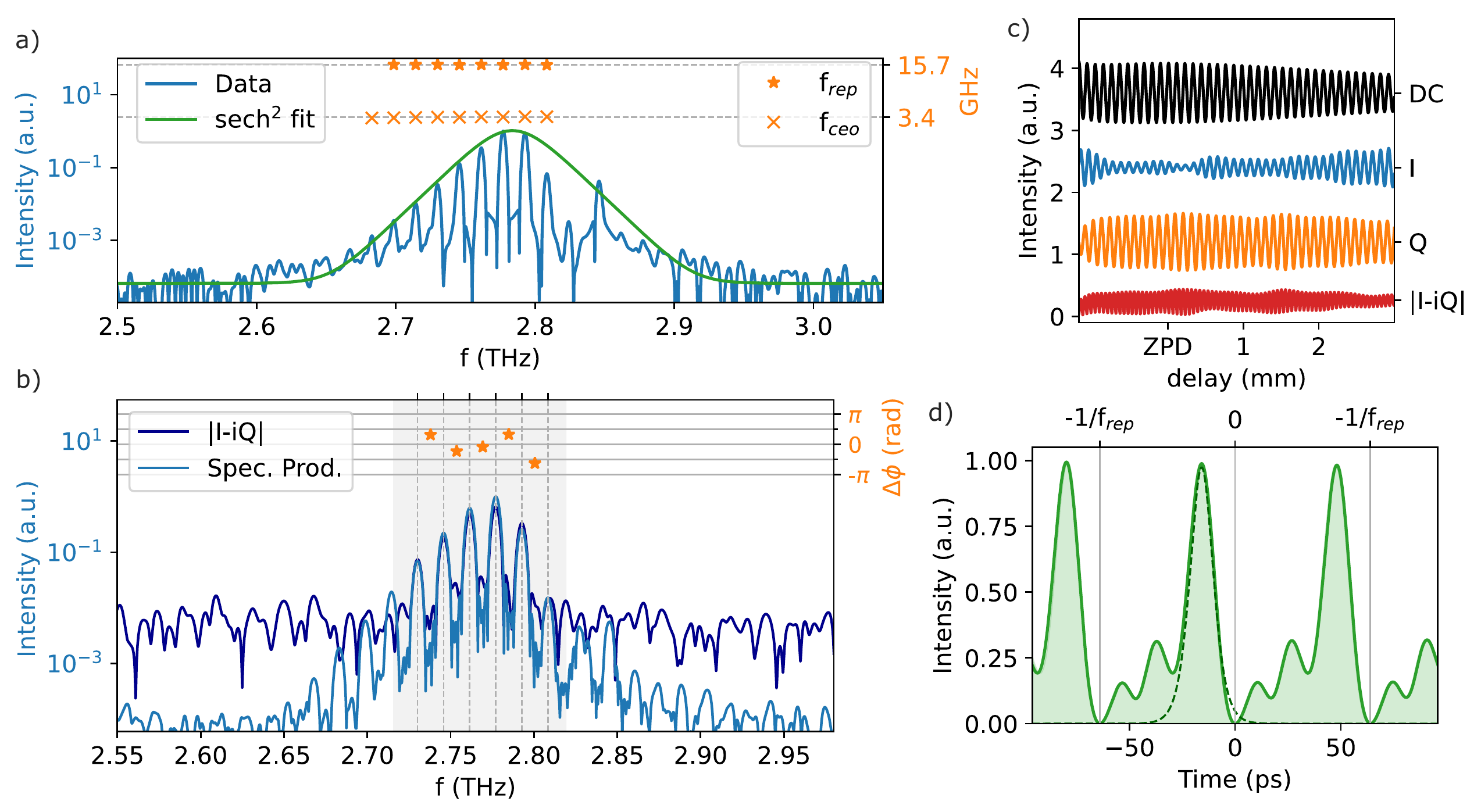}
	\caption{SWIFT spectroscopy of of a 800 $\mu$m radius, antenna-coupled, double ring laser. The laser is driven in CW at 500 mA (7.7V) and mildly RF injected with -15 dBm signal at 15.65 GHz. The DC spectrum (measured with the DTGS detector) is reported in a) and superimposed with a sech$^2$ fit of the envelope and the $f_{rep}$ and $f_{ceo}$ computed from the DC measurement. In the panel b) the spectral product (namely $I_{DC}(\omega) I_{DC}(\omega-f_{rep})$, light blue trace) and $|$I-iQ$|$ spectrum (measured with the HEB, dark blue trace) are displayed in a). The phase differences between the modes are shown on top to the spectra, in between the corresponding modes (orange $\star$). The grey area indicates the considered spectral region, i.e. where the signal of the HEB is above the noise floor. The corresponding reconstructed intensity profile is reported in b (solid line) together with a sech$^2$ fit of the pulse (dahsed line). The DC components and the IQ demodulated optical beatnote are shown in c). }
	\label{fig:SWIFT}
\end{figure*}

Thanks to the introduction of the bullseye antenna, that eases the characterization, the second generation of devices shows an overall highly improved behaviour. In particular we report a free-running device operating as a comb with narrow beatnote ($<$ 10 kHz in linewidth) and a clean sech$^2$ spectral envelope through almost its entire operation range (Fig \ref{fig:hyst} a-c). This laser also shows a pronounced hysteretic behaviour. A relatively fast variation (on a time scale $\lesssim$ ms) in the bias current or an RF injection is indeed needed to induce the comb state, which is robust against further variations of the bias current. On the contrary, if the laser bias is ramped-up slowly (on a time scale $ \gtrsim$ s), single mode operation can be observed in the entire operating range, until roll-over (Fig. \ref{fig:hyst} d). This hysteretic behaviour is even clearer if we inject an RF signal into the device bias line and we cycle up and down the injected power. In figure \ref{fig:hyst} e) we report the result of this experiment. The device is initially single mode: an RF tone, at the nominal repetition rate (15.7 GHz), is injected through a bias tee. As the RF power is increased, few sidemodes appear next to the main lasing mode until +15 dBm, where a sudden spectral broadening is observed. Decreasing the injected power the spectral bandwidth is slightly decreased, nevertheless the laser does not go back into the single mode operation but rather stays in a comb state featuring the typical sech$^2$ envelope observed in other devices. This behaviour is reflected into the beatnote intensity (Fig. \ref{fig:hyst} e), inset), which is dominated by the injected signal until -10 dBm (in the decreasing branch). There, the signal produced by the beating of the comb modes dominates over the external injection. This also suggests that the laser state is not strongly perturbed by the RF signal if the injected power is lower than -10 dBm, observation supported also by the low injection locking range ($\sim$ 10 MHz) at this RF intensity [S5]. 

The multimode operation seems to depend strongly on the GVD of the optical cavity. The laser emission is indeed within the spectral region where we expect anomalous dispersion, while devices emitting at higher frequencies, corresponding to positive GVD,  appear to be strictly single mode (Fig. \ref{fig:CGLE} a). The different behaviour of the two devices can be captured by the complex Ginzburg-Landau equation\cite{franckie2022self} which can be written, in the normalized form, as:
\begin{equation}
    \frac{\partial\mathcal{E}}{\partial\tau} = \mathcal{E} -(1-i\alpha)|\mathcal{E}|^2\mathcal{E} + (1+i\beta)\frac{\partial^2\mathcal{E}}{\partial\Theta^2}.
\end{equation}

Where $\mathcal{E}$, $\tau$ and $\Theta$ are the normalized electric field, time and spatial coordinate respectively, $\alpha$ is the negative of the linewidth enhancement factor (LEF) and $\beta = \delta/\epsilon$ is the ratio between the GVD ($\delta$) and the gain curvature ($\epsilon$) [see the Supplementary for details]. Depending on the values of $\alpha$ and $\beta$ the system can support DKS or evolve towards single mode or chaotic solutions. For the simulations presented in Figure \ref{fig:CGLE} d-e we considered an LEF of 1.1, while $\beta$ was found approximating $\epsilon$ with $g_0/\gamma_{trans}^2$, where the width of the laser transition $\gamma_{trans}$ $\sim$ 1 THz was determined with a measurement of the luminescence [S3]. The excess gain $g_0$ is evaluated experimentally using a ridge laser, based on the same active region, biased below threshold as a regenerative quantum detector\cite{micheletti2021regenerative}. A silicon lens is mounted on the facet of the device on which the emission of a similar laser is focused. In this way, the responsivity of the device used as detector is evaluated as a function of the emission frequency of the source laser for different bias currents of the detector (Fig. \ref{fig:CGLE} b). Fitting the experimental measurements of the responsivity it is possible to evaluate the gain $g_{exp}$ as a function of the current density J (Fig. \ref{fig:CGLE} c). The excess gain $g_0$= 3 cm$^{-1}$ is therefore found considering $g_{exp}$ at the current density relative to the spectra presented in Figure \ref{fig:CGLE} a ($J_{ring}$ = 149.2 A/cm$^2$ for both devices) and subtracting the waveguide ($\alpha_{w}$ = 6.8 cm$^{-1}$) and radiative losses ($\alpha_{rad}$ = 0.05 cm$^{-1}$) which are evaluated with numerical simulations [S3]. Assuming the GVD to be equal to the designed value at the central emission frequency, we could faithfully reproduce the state of the two lasers (Fig. \ref{fig:CGLE} d). The intensity distribution obtained in the case of multimode operation (Fig. \ref{fig:CGLE} e) shows a strong amplitude modulation on a non-zero background which matches well the reconstructed time trace observed in similar devices (Fig. \ref{fig:SWIFT} b). Shifted Wave Interference Fourier Transform Spectroscopy (SWIFT)\cite{burghoff2015evaluating} was indeed performed on a DR device with the same design and based on a similar active region, with a slightly less diagonal transition, to characterize the comb phases and reconstruct the time profile of the emission intensity. The set-up used to perform this characterization, described in detail in \cite{forrer2022spectral}, exploits a superconducting Hot Electron Bolometer (HEB)\cite{martini2021waveguide} to measure the comb beatnote which is then IQ demodulated with a spectrum analyzer. To do so the comb is RF injected using a source which is synced to the spectrum analyzer in order to keep the beatnote stable during the measurement and providing the IQ demodulation reference. Here we report a measurement performed with an injection power of -15 dBm, for which we can reasonably assume that the laser is very weakly perturbed by the RF signal. In Fig \ref{fig:SWIFT} a-b)  we report the DC spectrum measured with a DTGS and the relative spectral product which is superimposed with the one reconstructed from the HEB signal. The good overlap between the two spectra is a first indication of the state coherence. The lower signal-to-noise ratio of the HEB spectrum, on the other hand, can be attributed to higher bandwidth used in the measurement of the optical beatnote which leads to a noise floor of N$_{HEB}$ = NEP$\cdot\sqrt{f_{BW}}\sim$ 1 $\mu$W (with 800 pW/$\sqrt{Hz}$\cite{torrioli2022high} noise equivalent power (NEP) at 15 GHz and bandwidth of the IQ demodulator f$_{BW}$ = 2MHz ) in view of the $\sim$ 2 nW noise obtained for the DTGS. The reconstructed time profile shows the presence of pulses $\sim$ 12 ps in width over a weak background, thus indicating an amplitude modulated comb. This is in clear contrast to what observed in standard Fabry-Perot THz QCLs which behave as frequency modulated combs with a quasi-constant output, while showing pulsed operation only under strong injection ($\gtrsim$ 30 dBm)\cite{senica2022planarized, forrer2022spectral}. The phase difference between adjacent modes, reported on top of the spectra, is not perfectly flat. However it is possible to distinguish a region below $\sim$ 2.78 THz where the spectrum has a clear sech$^2$ envelope and the phases are dispersed around zero, and one above 2.78 THz where they strongly deviates from zero similarly to what observed in mid-IR ring QCLs\cite{meng2022dissipative}. This particular phase distribution is in stark contrast with the phase difference that characterizes frequency modulated combs, which were observed in devices whose emission is not centered onto the region with negative GVD, e.g. the single mode device presented in figure \ref{fig:CGLE} a). Under RF injection this device presents a comb spectrum with a $\sim$constant output in time and a linear phase difference between the modes spanning the whole interval from $\pi$ to -$\pi$ [S6]. While a moderate RF injection (+17 dBm in this case) can induce mode proliferation, indeed, it is the interplay between dispersion and non-linearity which appears to be the discriminant between amplitude and frequency modulated combs in double ring QCLs.


\section{Discussion and conclusions}

In this work we demonstrated the generation of DKS states in THz ring QCLs by means of dispersions engineering, leveraging on a fabrication technique based on planarization with BCB\cite{senica2022planarized}. Sech$^2$ shaped spectra are observed in defect-less double ring devices featuring negative GVD in free-running and under RF injection, hinting at the presence of DKS. The devices are further improved with the integration of a bullseye antenna which enhances the output power and far-field properties without introducing additional sources of back-scattering. On these devices comb operation with narrow beatnotes, clean sech$^2$ spectral envelopes and strongly hysteretic behaviour are observed indicating the presence of DKS. These observations are supported by SWIFT spectroscopy measurements of a ring under weak injection which confirms its operation as amplitude modulated self-starting comb with a pulse length of 12 ps. The experimental results are flanked by numerical simulations which reproduce the spectra and the intensity profiles observed experimentally. Beyond the fundamental interest on spontaneous soliton formation in THz QCLs, our approach is a promising way to obtain ultrashort THz pulses in free-running semiconductor lasers, thus constituting an extremely compact source for sensing and spectroscopy applications. Meanwhile the operating temperature of THz QCLs is approaching room temperature\cite{bosco2019thermoelectrically, khalatpour2021high}, our technological platform offers several ways to improve the present devices. A better power extraction can be achieved implementing bus waveguides and collecting more efficiently and eventually amplifying the in-plane scattered light. At the same time wider bandwidths could be obtained using more complex designs and optimizing the active region dispersion and non-linearity through band-structure engineering.



\subsection*{Supplementary Material}
See supplementary material for details about the device simulations and additional supporting measurements.
\subsection*{Data Availability} 
The data that support the findings of this study are available from the corresponding author upon reasonable request.
\subsection*{Acknowledgements} T
he authors gratefully acknowledge the financial support of the Swiss National Science Foundation (SNF) and from H2020 European Research Council Consolidator Grant (724344) (CHIC). 
\subsection*{Competing Interests} 
The authors declare that they have no competing financial interests.
\subsection*{Authors contributions} 
P.M., G.S. and J.F. conceived the idea. P.M. designed and fabricated the devices, carried out all the measurements, analysed experimental data and performed numerical simulations under the supervision of G.S. and J.F.. A.F. designed the RF PCB board and built the SWIFTS setup. U.S. developed the planarized waveguide fabrication process. S.C. and G.T. provided the HEB detectors, A.F. and G.S. optimized the HEB RF coupling. M.B. performed the epitaxial growth. M.F. provided the solver of the Ginzburg-Landau equation. P.M. and G.S. wrote the manuscript. All authors discussed the results and commented on the manuscript.
 
\subsection*{Correspondence}  
*Correspondence should be addressed to P. Micheletti (email: pmichletti@phys.ethz.ch) and G. Scalari (email: scalari@phys.ethz.ch).

\subsection*{Data availability} All the simulation and experimental data supporting this study are available from the corresponding author upon reasonable request.

\section*{References}\label{References}
\bibliography{bibtex-library.bib}

\end{document}